\documentstyle[aps,preprint]{revtex}
\tightenlines
\begin{document}
\title{Ferroelectrics with Low-energy Electronic Excitations }
\author{T.K. Ng}
\address{Hong Kong University of Science and Technology, Hong Kong}
\author{C.M. Varma}
\address{Bell Laboratories, Lucent Technologies, Murray Hill, NJ 07974}
\maketitle
\begin{abstract}
We formulate a general theory of the dielectric response of a lattice with
a structural transition and a polarisation instability due to
 a soft-optic mode coupled to low energy electronic excitations by the electromagnetic fields. The electronic excitations considered are in
 two-limits; those of a low density of free-electrons or those of a low
 density of strongly localised electrons in  the Coulomb-glass phase. The ferroelectric 
transition in the absence of the electronic-excitations and the low energy dielectric properties are shown to be strongly modified.
\pacs{}
\end{abstract}

\narrowtext
\section{ Introduction}
In the last few years there has been an enormous revival in the study of 
ferroelectric and ferroelectric materials \cite{lines},\cite{cross}, \cite{samara}. In a large part it is due to the investigation of complex 
materials with very large number of atoms per unit cell and substantial 
disorder. The disorder also introduces low-energy electronic states. Such materials display what has come to be known as `relaxor-ferroelectric'
properties. This term appears to cover all properties, not seen in simple ferroelectrics- for example $BaTiO_3$. Generally these materials do not have 
a ferroelectric transition but do have a very large dielectric-constant, the polarization displays complicated hysteresis loops and has a strong frequency and history dependence at low energies.

Various ideas for the form of disorder and its effect on the 
ferroelectric transition have been proposed 
\cite{Vugmeister},\cite{Viehland},\cite{blinc}. 
But to our knowledge, the important effect of 
low-energy particle-hole excitations due to shallow electronic levels has 
not been treated. Ferroelectricity (like superconductivity) is rather special because the order parameter, the polarization, couples to the low-energy 
electronic excitation through the electromagnetic-fields. In this paper we study such effects. The effects of disorder are neglected here by adopting
a virtual-crystal aproximation. In fact both effects, disorder and low-energy electronic excitations are important; we hope to treat them together in the future.

This paper is organised as follows. In Section 2, we present a general Lagrangian for a polarization instability coupled to electronic excitations
through the electromagnetic fields. The generalization to the disordered case is
also indicated. In section 3, we consider the simple case that the low energy excitations are due to a low density of free electrons. While it is common-place knowledge that free-electrons will screen the electric field so that macroscopic fields cannot arise, it is worthwhile to precisely derive the conditions for this and see the condition for the formation of ferroelectric domains at low enough charge density. The domain walls in this case may be charged. We also
derive the low energy charge and current response functions for this case.
In Sec. 4 we take up the more interesting problem of localised electrons in a Coulomb glass phase and derive the charge response functions.

\section{The Lagrangian for soft optic modes in
presence of low-energy electronic excitations}
   We consider the following Lagrangian density (at imaginary times), $L=
L_{ferro}+L_{EM}+L_{int}+L_{e}$, where
\begin{mathletters}
\label{lang}
\begin{equation}
\label{lferro}
L_{ferro}={1\over2\chi}\left[({\partial\vec{P}\over\partial\tau})^2
+\omega_0^2\vec{P}^2+\gamma(\nabla\vec{P})^2\right]+{1\over4}u
\vec{P}^4,
\end{equation}
is a Langrangian describing the polarization vector $\vec{P}$. For
$\omega_0^2=\alpha(T-T_c)$ Eq. (1a) describes an instability of the system
(in the absence of coupling to electric fields) when $T\rightarrow{T}_c$. 
\begin{equation}
\label{lem}
L_{EM}={1\over8\pi}\left[({1\over{c}}{\partial\vec{A}\over\partial\tau})^2
+(\nabla\phi)^2\right]+{1\over8\pi}(\nabla\times\vec{A})^2
\end{equation}
is the usual Lagrangian for EM fields in terms of vector potential
$\vec{A}$ and scalar potential $\phi$ in the Coulomb
 gauge, $\nabla.\vec{A}=0$. In the microscopic theory, $\omega_0$ is the frequency of the zone-center transverse optic phonon. The polarization vector $\vec{P}$ couples to
the electric field $\vec{E}$ through the usual Lagrangian $L_{int}$, where
\begin{equation}
\label{lint}
L_{int}=\vec{P}.\vec{E}=
-\vec{P}.\left[{1\over{c}}{\partial\vec{A}\over\partial\tau}+
\nabla\phi\right].
\end{equation}

  In the presence of charges, the EM potentials pick up 
additional terms in their Lagrangian. In the Coulomb gauge they are
\begin{equation}
\label{le}
L_{e}=-{1\over2}\left[\chi_L\phi^2+({1\over{c}})^2\chi_T\vec{A}^2\right],
\end{equation}
\end{mathletters}
where $\chi_L$ and $\chi_T$ are the longitudinal and transverse
response functions of the electrons  to scalar and vector potentials
$\phi$ and $\vec{A}$, respectively. In Eqs. (1), $\vec{P},\phi$ and 
$\vec{A}$ depend on
the position $\bf r$. For the translational invariant case, this may be directly
Fourier-transformed. In the case of disorder, they are random functions of
position and the response functions $\chi_L$ and $\chi_T$ are functions of 
two co-ordinates. In what follows, we shall assume an effective translational
invariant approximation, where disorder is considered averaged over at the outset.

   In the Coulomb gauge, the Lagrangian can be separated into 
longitudinal and transverse parts, $L=L_L+L_T+L_u$. We obtain after
fourier transforming,
\begin{mathletters}
\label{lf}
\begin{equation}
\label{ll}
L_L={1\over2}(v(q)^{-1}-\chi_L)\phi^2-{1\over2\chi}[(i\omega)^2-
(\omega_0^2+\gamma^2q^2)]\vec{P}_L.\vec{P}_L-i(\vec{P}_L.\vec{q})\phi,
\end{equation}
where $v(q)=4\pi/q^2$ is the Coulomb potential, and
\begin{equation}
\label{lt}
L_T=-{1\over8\pi{c}^2}[(i\omega)^2-c^2q^2+4\pi\chi_T]\vec{A}^2
-{1\over2\chi}[(i\omega)^2-(\omega_0^2+\gamma^2q^2)]\vec{P}_T.
\vec{P}_T-i\vec{P}_T.({i\omega\vec{A}\over{c}}),
\end{equation}
where we have divided the polarization vector $\vec{P}$ into longitudinal
and transverse parts $\vec{P}=\vec{P}_L+\vec{P}_T$, and
\begin{equation}
\label{lu}
L_u={u\over4}(\vec{P}_L+\vec{P}_T)^4,
\end{equation}
\end{mathletters}
is the fourth-order term in $L_{ferro}$. Notice that $L_u$ mixes the 
longitudinal and transverse parts of $\vec{P}$.

   To study the ferroelectric instability, we take the $i\omega=0$
limit and first minimize the free energy with respect to the $\phi$ and 
$\vec{A}$ fields. We obtain
\begin{eqnarray}
\label{pot}
\phi(q) & = & {i\vec{q}.\vec{P}_L\over{v}(q)^{-1}-\chi_L(q,0)} \\  \nonumber
{\vec{A}\over{c}} & = & -4\pi{i}{i\omega\vec{P}_T\over
(i\omega)^2-c^2q^2+4\pi\chi_T(q,0)}\rightarrow0.
\end{eqnarray}
Substituting these back into Eq. \ (\ref{lf}), we obtain an
effective free energy in terms of $\vec{P}$ only,
\begin{equation}
\label{fp}
f={1\over2\chi}(\omega_o^2+\gamma^2q^2)[(\vec{P}_L)^2+(\vec{P}_T)^2]
+{1\over2}{(\vec{q}.\vec{P}_L)^2\over[v(q)^{-1}-\chi_L]}
+{u\over4}\vec{P}^4.
\end{equation}
   In an infinite medium $P_T$ decouples from EM field completely in the
{\em static} limit, as is required by Faraday's Law that $\dot{\vec{B}}=
\nabla\times\vec{E}\neq0$ when a transverse electric field exists. This is not true in a finite sample, where a macroscopic $P_T$, if it exists is discontinuous at the surface. A surface charge - density $\sigma = \vec{P}.\hat{n}$ is then generated leading to an electric field.

\section{Free electrons}
For free electrons, we may use the fact that
\begin{equation}
(v(q)^{-1}-\chi_L(q.0))^{-1}\rightarrow{4\pi\over{q}^2+q_{sc}^2}  
\end{equation}
in the small $q$ limit. In Eq.(5), $q_{sc}$ is the static screening length, 
given in terms of density by the Thomas-Fermi length at low temperatures 
and by the Debye length in the classical regime at high temperatures, i.e.
\begin{equation}
q_{sc}^2\sim{q}_{TF}^2 = 6\pi n_0e^2/E_f = 4(3/\pi)^{1/3}n_0^{1/3}/a_0
\end{equation}
for $T<<E_f$. Here $n_0$ is the density of electrons, $E_f$ their Fermi-energy and $a_0$ the Bohr radius, and
\begin{equation}
q_{sc}^2\sim{q}_{D}^2 = 4\pi n_0e^2/kT.
\end{equation}
at high temperature. For most problems of interest in ferroelectrics, the
density is low enough that near room temperature the Debye Screening is applicable.

  In the absence of free electrons, an instability developes in $\vec{P}_T$
field in the $\vec{q}\rightarrow0$ limit as $T\rightarrow{T}_c$ (or
$\omega_o\rightarrow0$) always before any possible instability in $\vec{P}_L$.
This instability occurs at $T_c$, where the transverse optic phonon frequency
$\rightarrow 0$. The magnitude of $T_c$ will in general change in the
presence of free electrons due to microscopic effects not treated here. The
presence of free electrons bring the transition in $\vec{P}_T$ and 
$\vec{P}_L$ to the same temperature $T_c$ because of screening. i.e. in 
infinite samples, no uniform electric-field can be produced due to a 
macroscopic  $\vec{P}$ because the electric field sets up by the ions
oscillation is screened. The electric field in the ordered state of the 
system with $<\vec{P}_T>\neq0$ is given by
\[
\vec{E}=-\nabla<\phi>=\lim_{q\rightarrow0}\left({q^2 \sigma \over
{v}(q)^{-1}-\chi_L}\right).  \]
Here $\sigma$ is the surface charge density at the surface of the sample
at which $\vec{P}_T$ has a normal component $\vec{P}_T.\hat{n}\neq0$. 
$\vec{E}$ goes to zero as long as $\chi_L(q,0)$ is finite, i.e. 
$<\vec{P}_T>\neq0$ does not lead to uniform electric field
in the presence of free electrons. This does not necessarily mean that
ferroelectricity is destroyed by arbitrarily small density of electrons. 
As shown below a domain structure of electric-fields is still possible 
provided the charge density is low enough. 
 
  At temperatures $T<T_c$, it is easy to minimize the energy to show that
$<\vec{P}>=\vec{P}_o=\sqrt{{-\omega_o^2\over\chi{u}}}$. The Langrangian
can be written as $L=L(P_o)+L(P')$, where $\vec{P}'$ are fluctuations
around $\vec{P}=\vec{P}_o$. We find that to
second order in $P'$, $L(P')=L_L(P'_L)+L_T(P'_T)$, where
\begin{mathletters}
\label{lp'}
\begin{equation}
\label{lpl}
L_L(P'_L)=-{1\over2}\left[{1\over\chi}((i\omega)^2-(-2\omega_o^2+
\gamma^2q^2))-{4\pi{q}^2\over{q}^2+q^2_{sc}}\right](\vec{P}'_L)^2
\end{equation}
and
\begin{equation}
\label{lpt}
L_T(P'_T)=-{1\over2}\left[{1\over\chi}((i\omega)^2-(-2\omega_o^2+
\gamma^2q^2))-{4\pi(i\omega)^2\over(i\omega)^2-c^2q^2+4\pi\chi_T}\right]
(\vec{P}'_T)^2.
\end{equation}
\end{mathletters}
Notice that the only difference between $T>T_c$ and $T<T_c$ in the
quadratic fluctuations in $\vec{P}$ is that $\omega_o^2$ is replaced by
$-2\omega_o^2$ as temperature decreases from $T>T_c$ to $T<T_c$.

\section{domain wall formation}

  For usual (insulating) ferroelectrics with finite dimensions, domains are 
formed to minimize the electric field energy $\sim\int{d}^3x{\vec{E}^2\over
8\pi}$, coming from effective {\em surface} charges $\sigma(\vec{x})\sim
\hat{n}.\vec{P}(\vec{x})$, where $\hat{n}$ is a surface unit vector. In this
section the criteria of domain formation in the presence of free charges is
studied. For simplicity we shall consider a ferroelectric with a finite 
width $2L$ in $z$-direction, and with infinite extent in $x-y$ plane. We 
shall assume a transverse polarization field $\vec{P}_T$ with $\vec{P}_T$
pointing in $z$-direction, and with magnitude function of $x-$ and $y-$
coordinates, i.e. $\vec{P}=\hat{z}P(x,y)$ for $-L<z<L$. For a finite $L$ 
surface charges $\sigma(x,y)=(\delta(z-L)-\delta(z+L))P(x,y)$ will be set 
up, and the total energy of the system is given in momentum space by
\begin{eqnarray}
\label{ffinite}
F & = & \sum_{\vec{q}}\left[{1\over2}(v(q)^{-1}-\chi_L)|\phi(\vec{q})|^2-
\sigma(\vec{q})\phi(-\vec{q})
+{1\over2\chi}(\omega_0^2+\gamma^2q^2)|\vec{P}(\vec{q})|^2\right] \\  \nonumber
& & +{u\over4}\int^L_{-L}dz\int\int{d}xdy\vec{P}^4,
\end{eqnarray}
where $\sigma(\vec{q})=(2i)sin(q_zL)P(q_x,q_y)$ is the fourier transform of
the surface charge density. Notice that $\vec{q}.\vec{P}_T(\vec{q})=0$ in 
the sample. Minimizing the energy with respect to $\phi$ we obtain
\begin{eqnarray}
\label{ff1}
F & = & \sum_{\vec{q}}\left[{1\over2}{(2sin(q_zL)\
|P(q_x.q_y)|)^2\over{v}(q)^{-1}-\chi_L}+
{1\over2\chi}(\omega_o^2+\gamma^2q^2)|P(q_x,q_y)|^2\right]  \\  \nonumber
& & +{u\over4}\int^L_{-L}dx\int\int{d}xdy\vec{P}^4.
\end{eqnarray}
  Summing over $q_z$, and using the results that
\[
\int^{\infty}_{-\infty}dq_z{sin^2(q_zL)\over{q}_z^2+Q^2}
\sim\int^{\infty}_{-\infty}dq_z{1\over2(q_z^2+Q^2)}={\pi\over2Q},  \]
when $L\rightarrow\infty$, where $Q^2$ is a positive number, we obtain
\begin{eqnarray}
\label{fff}
F & \rightarrow & (L)\sum_{q_x,q_y}\left[{\pi\over{L}}
{\vec{P}(\vec{q})^2\over\sqrt{q^2+q_{sc}^2}}+{1\over\chi}
(\omega_o^2+\gamma^2q^2)\vec{P}(\vec{q})^2\right]  \\  \nonumber
& & +\int\int{d}xdy{uL\over2}\vec{P}^4,
\end{eqnarray}
where $\vec{q}=(q_x,q_y)$ in Eq.\ (\ref{fff}). Minimizing the free energy
with respect to $P(\vec{q})$ and $\vec{q}$ we see that the system has now 
an instability at finite $q$ which indicates the instability of the system 
to domain formation. Expanding the quadratic terms in $\vec{P}$ 
in Eq.\ (\ref{fff}) in powers of $q^2$ at small $q$, we find that the 
system has an instability towards domain formation when
\begin{equation}
\label{dc}
{\gamma^2\over\chi}-{2\pi\over{q}_{sc}^3L}<0,
\end{equation}
or when $q_{sc}\leq{L}_D^{-1}=({\pi\chi\over2\gamma^2L})^{1\over3}$. 
It is easy to show from Eq.\ (\ref{fff}) that in the absence of free
electrons ($q_{sc}=0$) $L_D^{-1}$ gives the characterestic size of 
the domains. The criteria for formation of domains \ (\ref{dc})
can thus be interpreted as simply saying that the screening length of
free electrons should be longer than the natural domain size in order 
for domains to be formed when free electrons are present.

The presence of a finite charge density has important consequences for the switching and hysteresis properties of domains by reversing external-fields.
Domain reversal must be accompanied now by motion of charges. The time scale
for this is now limited by the time for the charges to diffuse over the 
characterestic size of the domains.

\section{dielectric functions}
   To obtain the dielectric function at $T>T_c$, we neglect the $L_u$
term and integrate out the $\vec{P}$ fields first in the
Lagrangian \ (\ref{lf}). As a result, we obtain two effective
Lagrangain $L_{eff}(\phi)$ and $L_{eff}(\vec{A})$ for the electromagnetic
potentials. The dielectric functions can be obtained by identifying
\begin{eqnarray}
\label{ldi}
L_{eff}(\phi) & = & {\epsilon_L(q,i\omega)\over2v(q)}\phi^2,   \\  \nonumber
L_{eff}(\vec{A}) & = & {\epsilon_T(q,i\omega)[(i\omega)^2-c^2q^2]\over
8\pi}A_{\mu}^2,
\end{eqnarray}
where we have again considered the Coulomb gauge $\nabla.\vec{A}=0$.
After some algebra we obtain
\begin{mathletters}
\label{dielectric}
\begin{equation}
\label{dl}
\epsilon_L(q,i\omega)=1-v(q)\chi_L(q,i\omega)-{4\pi\chi\over(i\omega)^2-
\omega_o^2(q)},
\end{equation}
and
\begin{equation}
\label{dt}
\epsilon_T(q,i\omega)=1-{4\pi\chi_T(q,i\omega)\over(i\omega)^2}
-{4\pi\chi\over(i\omega)^2-\omega_o(q)^2},
\end{equation}
\end{mathletters}
where $\omega_o(q)^2=\omega_o^2+\gamma^2q^2$. A similar expression can also
be derived for $T<T_c$. The only difference is that $\omega_o^2$ is
replaced by $-2\omega_o^2$ when $T<T_c$. Notice that
$\omega_o(q)^2$ is positive definite at all temperatures $T\neq{T}_c$.

   The longitudinal electromagnetic collective mode dispersions
in the system can be obtained by solving the equation
$\epsilon_L(q,\omega(q))=0$. In the small $q$ limit and with
$\omega>>q$, $v(q)\chi_L(q,\omega)\rightarrow{\omega_P^2\over\omega^2}$,
where $\omega_P$ is the electronic plasma frequency, and we obtain
the secular equation, 
\begin{equation}
\label{lmode}
1-{\omega_P^2\over\omega^2}-{4\pi\chi\over\omega^2-\omega_o(q)^2}=0.
\end{equation}
(This equation is not modified in the classical limit for long-wavelengths.)
  Solving the equation we find that there are in general two solutions 
$\omega_1$ and $\omega_2$ in the
equation. At small $q$ and for $\omega_o(q)^2<<\omega_P^2$, we obtain
\begin{eqnarray}
\label{rootl}
\omega_1^2 & \rightarrow & \omega_P^2+4\pi\chi   \\  \nonumber
\omega_2^2 & \rightarrow & \left({\omega_P^2\over\omega_P^2+4\pi\chi}\right)
\omega_o(q)^2.
\end{eqnarray}
  The transverse electromagnetic collective mode dispersions are obtained
by solving $\omega^2\epsilon_T(q,\omega)-c^2q^2=0$. We obtain after some
simple algebra a secular equation very similar to Eq. \ (\ref{lmode}),
except that $\omega_P^2$ is replaced by $\omega_P^2+c^2q^2$, with
similar solutions
\begin{eqnarray}
\label{roott}
\omega_1^2 & \rightarrow & \omega_P^2+4\pi\chi+c^2q^2  \\  \nonumber
\omega_2^2 & \rightarrow & \left({\omega_P^2+c^2q^2\over\omega_P^2+c^2q^2
+4\pi\chi}\right)\omega_o(q)^2.
\end{eqnarray}
  Notice that $\omega_1$ gives the usual plasmon and polariton modes,
respectively for longitudinal and transverse EM fields in metals, with
modifications coming from polarization field, whereas $\omega_2$
gives the collective modes in longitudinal and transverse polarization fields
$\vec{P}_L$ and $\vec{P}_T$, respectively, with modifications coming
from metallic behaviour.Notice that at small $q$ and when
$\omega_o^2\rightarrow0$, $\omega_2$ becomes very small and will be in the
hydrodynamic regime where $\chi_L$ would have a diffusive form
\[
\chi_L(q,\omega)={dn\over{d}\mu}({Dq^2\over{i}\omega-Dq^2}). \]
In this regime, the dispersion $\omega_2$ will be modified. In 
particular, a finite damping of the collective mode will occur because 
of coupling to the electronic degrees of freedom.

\section{ferroelectrics with localized electrons}
  We shall consider here ferroelectrics in the presence of strongly
localized electrons in the Coulomb glass phase. In the dilute
particle-hole pairs approximation (see Appendix), the localized
electrons have only longitudinal response to external electric
fields, and with response functions $\chi_{CG}(q,\omega)$ given 
in the Appendix. Their effects on the ferroelectrics can be
studied by replacing $\chi_L$ for free electrons by $\chi_{CG}$.
First we consider $\omega=0$ and study ferroelectric instability 
and domain formation at different temperature ranges.

\subsection{ferroelectric instability and domain formation}
   As a function of temperature, the response function $\chi_{CG}(q,0)$
has three different regimes. First we consider high temperature 
$k_BT>>\Delta$ where the Coulomb glass effect is unimportant and the
response of the electrons are dominated by phonon-assisted hopping.
This is probably the regime of experimental interests in ferroelectrics. 
Using Eq.\ (\ref{chiph0}) for $Re\chi_{CG}(q,0)$ in Eq.(10) and repeating 
the analysis afterward we find that the free energy is given by Eq.(11)
with $\sqrt{q^2+q_{sc}^2}$ replaced by $q\sqrt{\epsilon}$ in first term
of Eq.(11), where $\epsilon\sim\epsilon_o+({4\pi^4g^2Aa_o^5\over1440})
({A\over{k}_BT})^{1\over4}$, where $\epsilon_o$ includes all the 
non-singular contributions to $\epsilon$. The characteristic domain 
size $L_D$ is 
given by 
\begin{equation}
\label{ld}
L_D\sim({2\gamma^2L\sqrt{\epsilon}\over\pi\chi})^{{1\over3}}, 
\end{equation}
which differs from the pure ferroelectric case only by the factor
$\epsilon$. Notice that $\epsilon$ is expected to be large in the
Coulomb or fermi glass regime and we expect that the effect of introducing
localized electrons in the system is to enlarge the domain size. Notice
that for small enough $L$, we may enter a regime where $L_D/a_o<<
(r_T/a_o)^{5\over2}$. In this regime, we should use Eq.\ (\ref{chiphq})
for $Re\chi_{CG}(q,0)$. In this regime, we find that the enhancement
in domain size by the factor $\epsilon$ is lost, and the analysis is
similar to the case of metals except that the screening length 
$q_{sc}^{-1}$ is given by
\[  q_{sc}^2\sim{4\pi^4e^2g^2\over144}a_o^3(k_BT).  \]
Notice that in contrast to metals, the screening length increases as
temperature is lowered.

  As temperature is lowered we find that for large enough $L$ the
domain size is still given by Eq.\ (\ref{ld}) except that the value
of $\epsilon$ goes thought two more regimes. At temperature $\Delta>
k_BT$ but $\omega_T>I(r_o)$, we find 
\[
\epsilon\sim\epsilon_o+{\pi{g}^2e^4a_o^4\over6}({T_o\over{T}})^2,  \]
whereas at $\omega_T<I(r_o)$, we find
\[
\epsilon\sim\epsilon_o+{n^2\over10\pi}ln({T_o\over{T}}).  \]
Notice that our theoretical prediction of domain size variation as a 
function of temperature can be tested experimentally. Note that for small 
enough $L$, the effect of screening will become stronger as in the high
temperature case $k_BT>\Delta$, and the enhancement factor $\epsilon$ in 
domain wall size is lost.

  It should also be kept in mind that the Coulomb glass has hysteretic low 
frequency response. One should therefore expect a slow and sluggish transition over a wide temperature range starting from about the $T_c$ to roughly when 
$\omega_0^2+4\pi \chi \approx 0$.

\subsection{(longitudinal) collective modes}
   At $T>T_c$, the longitudinal collective modes of the system can be 
obtained by solving 
\begin{equation}
\label{cgcol}
1-v(q)\chi_{CG}-{4\pi\chi\over\omega^2-\omega_o(q)^2}=0.
\end{equation}
   In the $\omega>>q\rightarrow0$ limit, we obtain the secular equation
\begin{equation}
\label{rootcg}
\omega^2=\omega_o(q)^2+{4\pi\chi\over1-v(q)\chi_{CG}(q\rightarrow0,\omega)},
\end{equation}
where $\chi_{CG}(q\rightarrow0,\omega)$ are given by Eqs.(A6a), (A6b),
at temperature range $k_BT<\Delta$ at $\omega<I(r_o)$ and $\omega>I(r_o)$,
respectively. At temperatures $k_BT>>\Delta$, $\chi_{CG}(q\rightarrow0,0)$
is given by Eq.(A9). Although the quantitative behaviours of the collective
modes are different, logarithmic corrections appear in the pure dielectric
collective mode frequency $\omega_o(q)$ in all the above cases. The 
logarithmic correction vanishes in the ferroelectric instability limit
$\omega\sim\omega_o(0)\rightarrow0$, in agreement with the static 
instability analysis which indicates that the ferroelectric instability 
occurs exactly when $\omega_o(0)\rightarrow0$. Notice, however that a 
collective mode solution with $\omega^2\sim\omega_o(q)^2+{4\pi\chi\over
\epsilon}$ exists even up to the instability limit 
$\omega_o(0)\rightarrow0$. The polarization collective modes are in 
general much broadened by the Coulomb glass particle-hole excitation 
spectrum because of the very slow vanishing of $Im\chi_{CG}(q,\omega)$
at small $\omega$. Notice that at $T<T_c$, we may obtain the collective
modes by replacing $\omega_o^2$ by $-2\omega_o^2$ in Eq.\ (\ref{cgcol}).
This is valid as long as we are interested at responses of the system at
wave vectors $q<L_D^{-1}$. 

\appendix

\section{Electromagnetic response of Coulomb glass}
   In this appendix we summarize our theoretical understanding for
response functions of Coulomb glass systems. (Notice that we have set
$e^2=1$ in the main text. Here $e^2$ is kept to make the equations
more clear). The electromagnetic responses of Coulomb glass systems
are rather complicated, and can be divided into many different 
regimes, depending on temperature, frequency and wavevector. A
detailed discussion of these can be found in ref[\cite{bhatt}].   

   The starting model for responses of Coulomb glass systems is the
effective two-site Hamiltonian by Shklovskii and Efros\cite{efros}. The Hamiltonian is derived under the assumption that the system is in the 
strongly  disordered regime where the electrons are strongly localized. 
In this regime tunneling events of electrons from one site to another is
rare and can be treated in a dilute particle-hole pairs approximation
where interaction between different tunneling events are neglected
and the tunneling of electron between two sites (particle-hole pair)
is described by a two-site Hamiltonian
\begin{mathletters}
\label{h2}
\begin{equation}
\label{tsites}
H_{ij}=E_in_i+E_jn_j+{e^2\over{r}_{ij}}n_in_j+I(r_{ij})(a^+_ia_j+
a^+_ja_i),
\end{equation}
where $n_i$ is the electron occupation number ($=0,1$) at site $i$,
$r_{ij}$ is the distance between the two sites, $I(r)\sim{I}_oe^{-r
/a_o}$ is the effective tunneling matrix element between the two
sites, where $a_o\sim$ lattice spacing and $a(a^+)$ are electron
annihilation and creation operators. the on-site energy $E_i$ is
given by
\begin{equation}
\label{eon}
E_i=\epsilon_i+\sum_{l\neq{i},j}{n_le^2\over|r_l-r_i|},
\end{equation}
\end{mathletters}
where $-A<\epsilon_i<A$ is the random on-site energy with bandwidth
$2A$ and the second term
comes from Coulomb interaction. Notice that electron spins are ignored
in the theory. $H_{ij}$ can be diagonalized easily to obtain the
following energy spectrum,
\begin{eqnarray}
\label{spectrum}
E_1^{\pm} & = & {1\over2}(E_i+E_j)\pm{\Gamma\over2}  \\  \nonumber
E_2 & = & E_i+E_j+{e^2\over{r}_{ij}}
\end{eqnarray}
where $E_1^+$ and $E_1^-$ are the two possible energy states when the
two-site system is occupied by a single electron, and $E_2$ is the energy
when both sites are occupied, $\Gamma=\sqrt{(E_i-E_j)^2+4I(r_{ij})^2}$.
Notice that for the pair of sites to be occupied by a single electron
and participate in the electromagnetic response, we must have $E_1^-<\mu<
E_2-E_1^-$, where $\mu$ is the chemical potential. The whole system
is described as a diluted gas of independent (i.e. non-interacting),
randomly distributed particle-hole pairs and the total response
of the system to external electromagnetic fields is given by
by sum of the response of individual pairs. For example, the
density-density response function of a pair in the two-site
Hamiltonian approximation is found to be (at temperature $T=0$)
\begin{mathletters}
\label{chicg}
\begin{equation}
\label{chipair}
\chi_p(\vec{q},i\omega)=2{I(r_{ij})^2\over\Gamma(r_{ij})^2}(1-cos(\vec{q}.
\vec{r}_{ij}))\left[{1\over{i}\omega-\Gamma(r_{ij})}-
{1\over{i}\omega+\Gamma(r_{ij})}
\right],
\end{equation}
and the average density-density response function of the whole system is
\begin{eqnarray}
\label{chiave}
\chi_{CG}(\vec{q},i\omega) & = & <\sum_{p}\chi_p(\vec{q},i\omega)>
\\  \nonumber
& \sim & 2\int^A_{-A}dE_1\int^A_{E_1}dE_2\int{d}^3r{I(r)^2\over
\Gamma(r)^2}(1-cos(\vec{q}.\vec{r}))F(E_1,E_2,r){2\Gamma(r)
\over(i\omega)^2-\Gamma(r)^2},
\end{eqnarray} 
\end{mathletters}
where $F(E_1,E_2,r)$ is
the probability that a particle-hole pair with on-site energies
$E_1$ and $E_2$ is found separated by distance $r$. For $E_1,E_2$
larger than the Coulomb gap energy $\Delta\sim(e^2/a_o)^{3/2}/A^{1/2}$,
$F\sim{g}^2$ where $g=n/Aa_o^3$ is the density of states above the
Coulomb gap, $n$ is the average number of electrons per site. For 
$E_1,E_2$ much
smaller than $\Delta$, $F\sim\rho(E_1)\rho(E_2)$, where
\[
\rho(E)={3n\over\pi{e}^6}E^2.  \]
Notice that a two-site system behaves as a electric dipole and has
only longitudinal response to external electric fields.
Evaluating the integrals, we find for $I(r_o)<<\omega<<A$,
\begin{mathletters}
\label{chii}
\begin{equation}
\label{chii1}
Im\chi_{CG}(q,\omega)\sim{a}_o\pi{g}^2e^2r_{\omega}(1-{sin(qr_{\omega})
\over{q}r_{\omega}}),
\end{equation}
where $r_{\omega}=a_oln({2I_o\over\omega})$, $r_o=e^2/\Delta$. 
For $\omega<<I(r_o)$,
we obtain
\begin{equation}
\label{chii2}
Im\chi_{CG}(q,\omega)\sim{3n^2a_o\over10\pi{e}^2}{1\over{r}_{\omega}^3}
(1-{sin(qr_{\omega})\over{q}r_{\omega}}).
\end{equation}
\end{mathletters}
  The real part of the response function can be obtained using the
Kramers-Kronig relation, we obtain for $qr_o<<1$,
\begin{mathletters}
\label{chiq}
\begin{equation}
\label{rechi0}
Re\chi_{CG}(q,0)\sim{n^2\over20\pi^2e^2}q^2ln(qr_o),
\end{equation}
and for $qr_o>>1$,
\begin{equation}
\label{rechi1}
 Re\chi_{CG}(q,0)\sim-{g^2\over2}e^2r_o^2=-{g^2\over2}r_o^3\Delta.
\end{equation}
\end{mathletters}
   For $I(r_o)>\omega$ and $qr_{\omega}<<1$, we obtain
\begin{mathletters}
\label{chiw}
\begin{equation}
\label{chire}
Re\chi_{CG}(q,\omega)\sim{n^2\over20\pi^2e^2}q^2ln(ln({\omega\over2I_o})),
\end{equation}
and for $\omega>I(r_o)$,
\begin{equation}
\label{chire1}
Re\chi_{CG}(q,\omega)\sim-{g^2e^2\over24}q^2r_{\omega}^4.
\end{equation}
\end{mathletters}

    At finite temperatures $T<\Delta$, eq.\ (\ref{chii2}) for $Im\chi_{CG}
(0,\omega)$ is cutoff at a low frequency $\omega_T\sim{e}^{-(T_o/T)^{1\over2}}$ 
by the thermal mechanism of Mott-variable range hopping\cite{bhatt} 
(in the presence of Coulomb gap) where $T_o\sim{e}^2/a_o$. Correspondingly,
at temperature $\omega_T<I(r_o)$ $Re\chi_{CG}(q,0)$ is replaced by\cite{bhatt}
\begin{equation}
Re\chi_{CG}(q,0)\sim{n^2\over20\pi^2e^2}q^2ln({a_o\over{r}_{\omega_T}})\sim
{n^2\over40\pi^2e^2}q^2ln({T\over{T}_o}),
\end{equation}
at $\omega<\omega_T, (q=0)$ and at $qr_o<r_o/r_{\omega_T}, (\omega=0)$. At
temperature $\omega_T>I(r_o)$, a corresponding analysis gives
\begin{equation}
Re\chi_{CG}(0,0)\sim-{g^2e^2\over24}q^2r_{\omega_T}^4\sim-{g^2e^2a_o^4\over24}
q^2({T_o\over{T}})^2,
\end{equation}
for the same low frequency range and at small wavevector range $qr_o<
(r_o/r_{\omega_T})^2$.

  At still higher temperatures $k_BT>>\Delta$ where the Coulomb gap 
becomes unimportant and phonon-assisted hopping becomes dominant, 
we enter another regime where $\chi_{CG}$ become different 
again\cite{bhatt}. At frequencies $qr_{\omega}<<1$ and 
$\omega>\bar{\omega}_T$, where $\bar{\omega}_T\sim\omega_P
e^{-(A/k_BT)^{1\over4}}$, where $\omega_P$ is a characteristic
phonon frequency, we obtain
\begin{equation}
\label{chiphw}
Re\chi_{CG}(q,\omega)\sim{\pi^3g^2k_BT\over1440}q^2r_{\omega}^5,
\end{equation}
and for $\omega=0$,
\begin{equation}
\label{chiphq}
Re\chi_{CG}(q,0)\sim-{\pi^3g^2\over144}a_o^3(k_BT).
\end{equation} 
  For frequencies $\omega<\bar{\omega}_T$ and for wavevectors
$qa_o<<(a_o/r_T)^{5\over2}$, where $r_T\sim({A\over{k_BT}})^{1\over4}a_o$ 
is the characteristic thermal hopping distance, $Re\chi_{CG}$ saturates 
at a value\cite{bhatt}
\begin{equation}
\label{chiph0}
Re\chi_{CG}(q,0)\sim-{\pi^3g^2k_BT\over1440}q^2a_o^5
({A\over{k_BT}})^{5\over4}.
\end{equation}  
Notice that the Mott variable range hopping introduce characteristic
cutoff length scales $r_{\omega_T})$ and $r_T$ at different temperature 
regimes. As a result the dielectric constant $\epsilon(q\rightarrow,0)=
1-v(q)\chi_{CG}(q\rightarrow0,0)$ becomes finite but large at low 
temperatures although the temperature dependence of its value goes 
through several different regimes\cite{bhatt}.

\end{document}